\documentclass[a4paper, oneside, twocolumn, notitlepage, 10pt]{extarticle_ecoc}
\usepackage{ecoc}

\begin{document}
\selectlanguage{english}    % Standard Language

%-------------------------------------------------- Title -----------------------------------------------------%

\title{Preventing Data Hallucinations in Agentic AI for Optical Network Troubleshooting}%

%------------------------------------------------- Authors-----------------------------------------------------%

\author{
    Jeremias D\"otterl\textsuperscript{(1)},
    Martin Kuipers\textsuperscript{(2)}
}

\maketitle                  % Create title and author

%------------------------------------------ Description of Authors ----------------------------------------------%

\begin{strip}
    \begin{author_descr}

        \textsuperscript{(1)} Adtran GmbH, Hermann-Dorner-Allee 91, 12489 Berlin, Germany,
        \textcolor{blue}{\uline{jeremias.doetterl@adtran.com}}

        \textsuperscript{(2)} Adtran Networks SE, Hermann-Dorner-Allee 91, 12489 Berlin, Germany,
        \textcolor{blue}{\uline{martin.kuipers@adtran.com}}

    \end{author_descr}
\end{strip}

% \setstretch{1.1}
%-------------------------------------------------- Footnote -------------------------------------------------------%
\renewcommand\footnotemark{}
\renewcommand\footnoterule{}
%\let\thefootnote\relax\footnotetext{text}

%-------------------------------------------------- Abstract ---------------------------------------------------------%

\begin{strip}
    \begin{ecoc_abstract}
        An agentic AI system for automated troubleshooting in optical communication networks is demonstrated that integrates data references into the agent architecture to prevent data hallucinations, improving the interpretability and trustworthiness of root-cause identification and reasoning. ©2026 The Author(s) 
    \end{ecoc_abstract}
\end{strip}

\section{Overview}

Service providers have a strong interest in automating root cause analysis (RCA) in Passive Optical Networks (PON) to minimize the impact of service degradations and outages.
Root cause models for PONs analyze telemetry acquired from the equipment, such as Optical Line Terminals (OLTs) and Optical Network Units (ONUs), and predict probable causes, such as fiber bends, loose or dirty fibers, laser aging, power outages, or rogue ONUs.
For use in real deployments, these models must be accurate and explainable to ensure reliability and operational trust.
Only if the claimed root cause is sufficiently justified by the data, operators can trust the inference and confidently resolve the issue with appropriate remedies.

Through recent advances in large language models (LLMs)~\cite{achiam2023gpt}, AI agents~\cite{acharya2025} have become capable of advanced language understanding and sophisticated reasoning.
Therefore, AI agents are a promising paradigm for automating the troubleshooting and root cause inference in optical networks to an unprecedented degree.
In a typical setup, the agent is given access to the PON telemetry, a knowledge base, and a set of tools, and is instructed via a prompt to analyze the PON telemetry and infer the most likely root cause.
However, while today's LLM-based agents are useful tools for assisting human experts in ad-hoc root cause analysis, their black-box nature~\cite{ajwani2024,deCarvalho2025} and hallucinations~\cite{bang2025,huang2025} prevent the adoption for wide deployments in commercial networks, where operators expect explainable inference and interpretable root cause claims.

One of the major challenges with today's AI agents is the risk of \emph{data hallucinations}.
Agents regularly generate responses that appear plausible but contain incorrect information \cite{huang2025}.
By default, agent responses consist of raw text, in which the agent outlines its root cause claims and (if prompted to do so) the reasoning chain that substantiates this claim.
However, these textual responses lack sufficient rigor that would allow operators to trust these responses.
Whenever the agent makes a statement about the underlying PON telemetry, there is a risk that the agent hallucinates data, i.e., it refers to data in its responses that is not present in the actual PON telemetry.
This makes standard AI agents unreliable and untrustworthy for troubleshooting commercial PONs.

In this demonstration, we present a new evidence-based framework for agentic network troubleshooting, in which the agent uses \emph{data references} to point to data fragments in the PON telemetry.
Using data references, the agent can cite specific subsets of the data in its claims and reasoning justifications, for example a spike or drop in a particular metric in a particular time window.
We show how data references can be integrated into the agentic AI architecture and demonstrate their benefits in various degradation and fault scenarios on commercial PON equipment.

\section{Novelty}
Currently, there is great interest in agentic AI methods for the monitoring, configuration and fault management of (optical) communication networks \cite{shi2025,Zaid:26,Arunachalam:26,vde:456426068} driven by fast progress in LLMs and agentic AI research.
Our demonstration is aligned with these developments and enhances the current agentic AI architecture.
The problem of hallucinations and low reliability of agentic AI (and generative AI more broadly) is a major challenge \cite{Arunachalam:26} that must be addressed.

The major novelty of the presented approach is the introduction of data references into the standard agentic AI architecture to enable evidence-based root cause analysis in optical networks.
Agentic consumer applications, like ChatGPT, use citations to link to information found on a web site.
Data references in our architecture serve a similar purpose, but they link to data sources and are tailored to the requirements in optical network troubleshooting.
Specifically, data references are designed to work with common telemetry data formats, such as time series data, log files, and structured JSON.
A data reference is a fine-grained pointer that not only identifies the data source but also a specific data fragment, such as a line range in a log file, a snippet in a JSON object, or a window of a time series metric.
In our implementation, data references are stored as URIs where parameters are used to encode information about the OLT, ONU, metric, time window, etc.

\begin{figure*}
    \centering
    \includegraphics[width=0.99\linewidth]{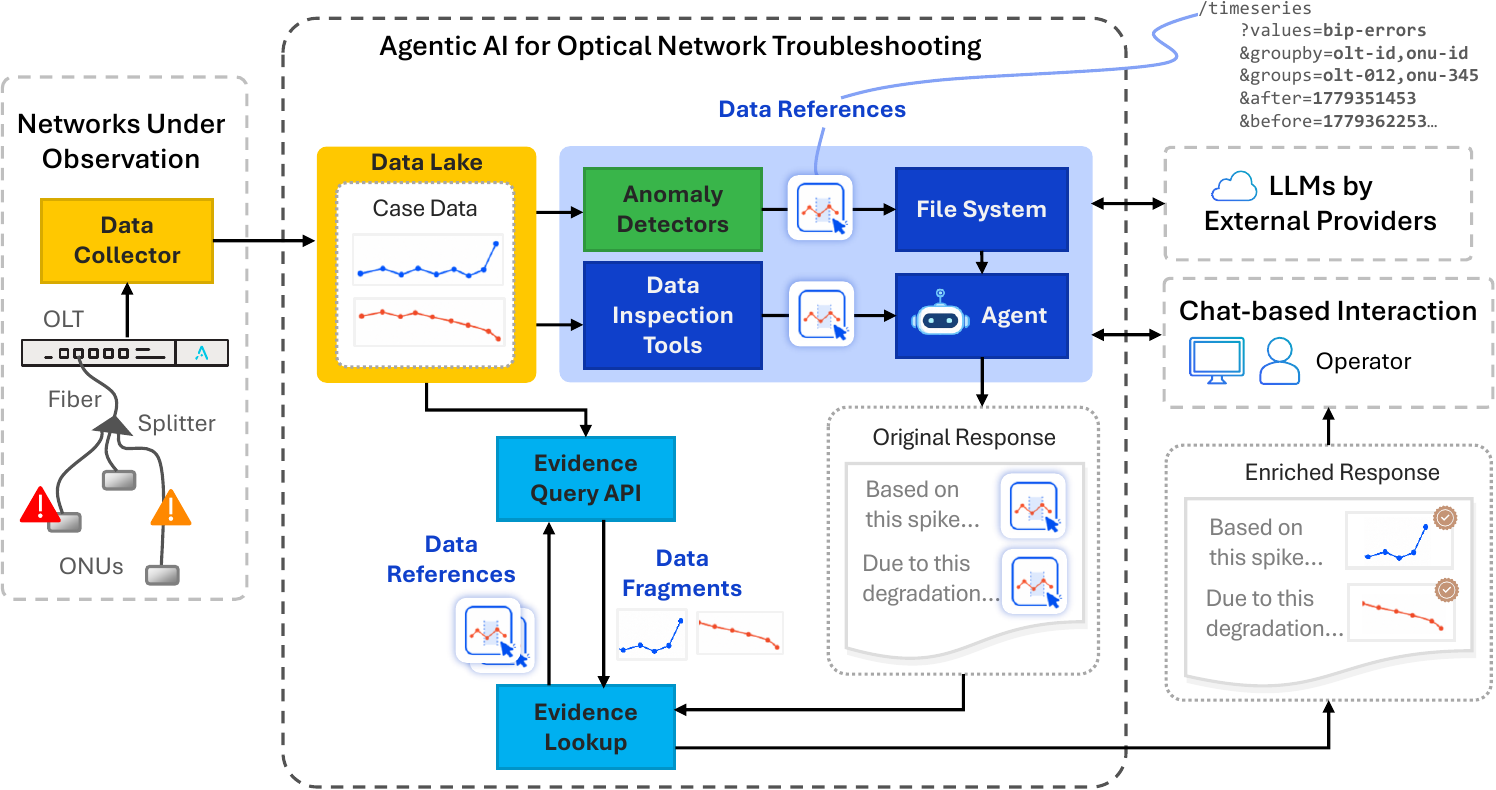}
    \caption{Agentic AI for Optical Network Troubleshooting}
    \label{fig:main}
\end{figure*}

Fig.~\ref{fig:main} shows the \emph{Agentic AI for Optical Network Troubleshooting} and its interaction with external systems and actors.
The \emph{networks under observation} are PONs that are monitored by a data collector.
Via the OLT, the data collector acquires telemetry that provides insights into the state and performance of the optical network, including traffic counters, bit-interleaved-parity (BIP) error counters, TX/RX power levels, as well as detailed alarm logs.
The networks can suffer impairments caused by fiber bends, loose connectors, sudden power outages, etc., whose diagnosis is time-consuming and costly in large installations.
Whenever the service quality degrades, the operator can start a \emph{chat-based interaction} with the agent to trigger the data collection and diagnosis process.
The data collector takes a snapshot of the recent OLT telemetry and uploads it to the \emph{data lake} in the cloud.
This \emph{case data} serves as the input data for the agent's diagnosis.

In a first analysis step, \emph{anomaly detectors} process the case data and identify anomalous data that deviates from expected data patterns.
These anomaly detectors can be implemented with heuristic rules or machine learning models.
Their purpose is to inject domain knowledge into the RCA process and to steer the agent's attention towards interesting data sections.
While this component is optional, it can help reduce the agent's token consumption and costs.
The anomaly detectors use data references to point to the data fragments where the anomaly was found.
The data references are written to a file system, where they can be inspected by the agent with standard tools like ls, grep, and read\_file.

The \emph{agent} uses an \emph{LLM by an external provider} for analysis and reasoning.
Guided by the system prompt, the agent inspects the data references in the file system.
Additionally, the agent is equipped with \emph{data inspection tools} that allow the agent to run data queries.
For example, the agent may use a tool to identify the temporal overlap of two anomalies, or to compute the correlation between two performance metrics.

One important technical innovation in this architecture is the enrichment of the agent--tool API contract:
All data inspection tools include data references in their responses.
This way, as the agent gathers information about the case data, it keeps track of the data locations.

\begin{figure*}
    \centering
    \includegraphics[width=0.86\linewidth]{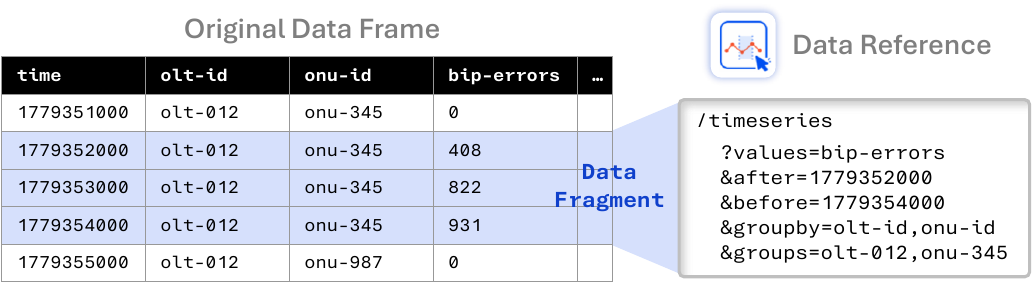}
    \caption{Data References}
    \label{fig:second}
\end{figure*}
When the agent generates its answer, it is not allowed to generate a copy of the data in its response, as this poses a risk of data hallucination.
Instead, the agent is instructed via the prompt to use data references to substantiate its claims.

Fig.~\ref{fig:second} shows an example:
The time series of BIP errors of each ONU is stored in a data frame.
If the agent wants to refer to an observed spike in the BIP error metric of ONU \emph{onu-345} at OLT \emph{olt-012}, it must include the correct data reference into its response.
In this URI, the \emph{values} parameter indicates the column of interest, the \emph{after} and \emph{before} parameters define the time window, \emph{groupby} defines the grouping columns and \emph{groups} defines a filter on these columns.
Similar URI schemas exist in our demo implementation to refer to line ranges of log files and snippets of JSON objects.

As shown in Fig.~\ref{fig:main}, the agent's \emph{original response} is sent to the \emph{evidence lookup} module, which replaces the data references with the actual data fragments from the case data.
We have developed an \emph{evidence query API} that accepts data references and extracts the correct data fragments from the original case data.
This enforcement mechanism guaranties that all the data that are displayed in the \emph{enriched response} exist in the case data.
If the agent hallucinated data references (which we have not observed in our experiments), the evidence lookup would fail, and a warning would be displayed to the user.
Finally, the enriched response is sent to the user interface where it is displayed to the user.

\section{ECOC Relevance}
Our demonstration contributes to the ECOC Demo Zone topic of \emph{AI/ML for network intelligence}.
The application of LLMs and agentic AI to optical networks is an emerging and rapidly evolving research area of significant relevance to the ECOC community.
At the same time, data hallucinations remain a major obstacle to the adoption of AI agents in operational networks, where incorrect decisions can have substantial operational and economic impact.
Consequently, improving the robustness, reliability, interpretability, and trustworthiness of AI agents is a critical requirement for delegating operational decisions to such systems.
In this context, our demonstration presents a novel integration of LLM-based agent systems and data references for optical network monitoring and troubleshooting.
The demo illustrates how combining the adaptability of modern AI agents with the explainability provided by data references can enable more trustworthy and resilient network operations.
As such, the work is relevant to both researchers and network operators attending ECOC.

\section{Demonstration Content and Implementation}
The objective of the demo is to demonstrate how an agentic AI system for troubleshooting optical network issues can be made reliable and trustworthy.
In particular, the demo illustrates how the proposed data-reference approach prevents data hallucinations, thereby improving the interpretability and trustworthiness of the root-cause claims generated by the agent.

The agentic AI system is implemented as a Python application using the \emph{deepagents}\footnote{https://github.com/langchain-ai/deepagents (Accessed: 2026-05-22)} library by \emph{LangChain} to realize the agent harness.
The agent operates in a reasoning loop in which it is guided by a prompt and invokes tools as needed.
Specialized tools were implemented that allow the agent to inspect the case data.
Both anomaly detectors and data inspection tools include data references in their responses.
Rendering functions were implemented to visualize these references with plots and quotes from the underlying telemetry data.
Most commercially available LLMs can be used within the system through LangChain's integrations with common model providers.

The demo involves multiple lab installations at Adtran premises, consisting of OLTs and multiple ONUs.
Similar to the techniques used by \cite{Sica:26,Sica2026}, representative faults will be provoked on these installations, including distribution fiber cuts, feeder fiber cuts, and sudden ONU power outages.
Through the data collector, the agentic system acquires live telemetry data from these installations.
This setup enables the demonstration of different failure modes and provides diverse troubleshooting scenarios.

Attendees will interact with the system through the web-based user interface.
They will be able to engage with the agent in multi-turn conversations.
The agent performs root-cause analysis (RCA) on live telemetry data and uses data references to point to relevant fragments of the underlying data.
The user interface visualizes the referenced PON telemetry through plots and quotes.
By contrasting this evidence-based implementation with an agentic analysis approach that does not use data references, the advantages of the proposed method for interpretability and trustworthiness are highlighted.

\clearpage
\section{Acknowledgements}
Partially funded by the German Federal Ministry of Research, Technology and Space (BMFTR) in the SUSTAINET-Advance project (Grant 16KIS2271K).

%-------------------------------------------------- Bibliography Section -------------------------------------------------------%
\printbibliography

\vspace{-4mm}

%%%%%%%%%%%%%%%%%%%%%%%%%%%%%%%%%%%%%%%%%%%%%
%---------------------------------------------- End of Document -----------------------------------------------%
\end{document}